# Wireless Fidelity Real Time Security System

V.C.K.P Arul Oli[1]
Assistant professor Dept. of Computer Application's
Dhanalakshmi College of Engineering, Chennai.
vckparuloli@yahoo.co.in
Elayaraja Ponram[2],
Consultant Ciber, Hyderabad
eponram@ciber.com

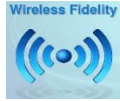

**ABSTRACT**

**This paper describes about how you can secure your Wireless Network from hackers about various threats to wireless networks, How hackers makes most use of it and what are the security steps one should take to avoid becoming victim of such attacks. There are plenty of opportunities to connect to public Wi-Fi hotspots when you're on the go these days. Coffee shops, hotels, restaurants and airports are just some of the places where you can jump online, but often these networks are open and not secure. Whether you're using a laptop, tablet or smartphone, you'll want to connect your device securely to protect your data as much as possible. Otherwise an unsecured Wi-Fi connection makes it easier for hackers to access your private files and information, and it allows strangers to use your internet connection. Here are some simple steps you can take to help make sure your data is safe on open public Wi-Fi.**

*Keywords* **– WEP, WAP, Wi – Fi, Wireless Fidelity Security, Encryption, Wireless Detection System**

## I. INTRODUCTION

This article describes how you can secure your Wireless Network from hackers and you'll also learn about free tools that people generally use to intercept your Wi-Fi signals. Wireless Networking (Wi-Fi) has made it so easy for you to use the computer, portable media player, mobile phones, video game consoles, and other wireless devices anywhere in the house without the clutter of cables.

With traditional wired networks, it is extremely difficult for someone to steal your bandwidth but the big problem with wireless signals is that others can access the Internet using your broadband connection even while they are in a neighboring building or sitting in a car that's parked outside your apartment.

This practice, also known as piggybacking, is bad for three reasons:

It will increase your monthly Internet bill especially when you have to pay per byte of data transfer.

It will decrease your Internet access speed since you are now sharing the same internet connection with other users.

It can create a security hazard* as others may hack your computers and access your personal files through your own wireless network.

**II. Who is connected to your Wireless Network?**

If you are worried that an outsider may be connecting to the Internet using your Wireless network, try Air Snare – it's a free utility that will look for unexpected MAC addresses on your Wireless network as well as to DHCP requests. Another option is that you open your router's administration page (using the 192.168.* address) and look for the DHCP Clients Table (it's under Status > Local Network on Linksys routers). Here you will see a list of all computers and wireless devices that are connected to your home network.

*It is also a good idea to turn off the router completely when you are not planning to use the computer for a longer period (like when you are out shopping). You save on electricity and the door remains 100% shut for wireless piggy backers.

**If you ever want to let a new device connect to your network, you will have to find its MAC address and add it to your router. If you simple want to let a friend connect to your wireless network one time, you can remove his MAC address from the router settings when he or she leaves your place.

**III. WIRELESS SECURITY BEST PRACTICES** 

How to Secure Your Wireless Network

The good news is that it is not very hard to make your wireless network secure, which will both prevent others from stealing your internet and will also prevent hackers from taking control of your computers through your own wireless network.

Here a few simple things that you should to secure your wireless network:

**Turn off all sharing**

To get to the network settings you'll need to change, open Network and Sharing Center from Control Panel. Then click "Change private advanced sharing settings" located in the left pane. Then under Private, Files and printer sharing, Guest or Public, turn off network discovery. Make sure to click Save changes for them to take effect. Also turn off file and printer sharing.

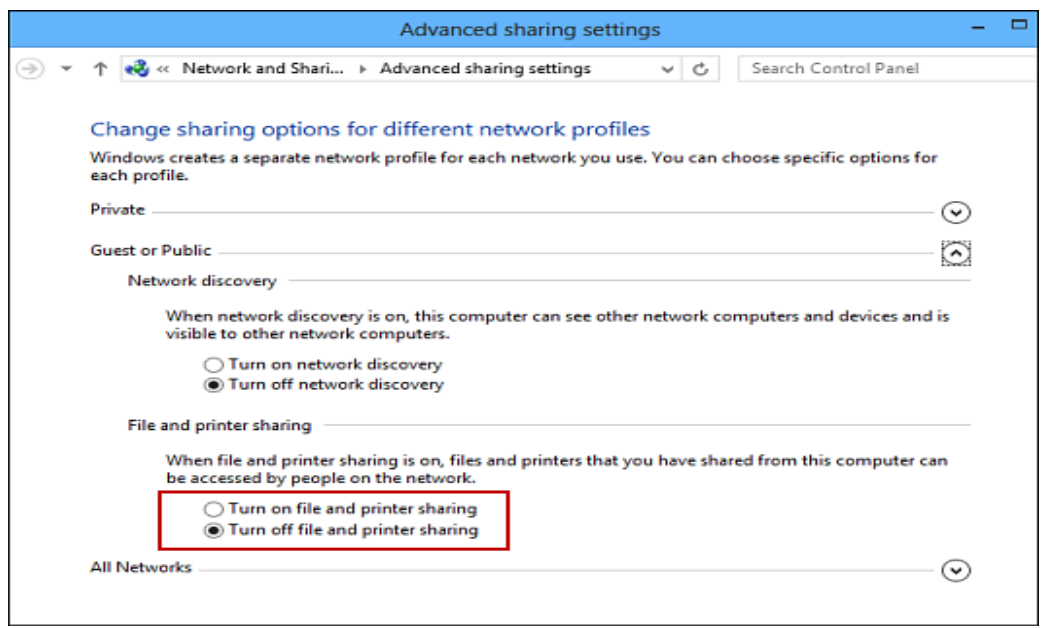

**Open Your Router Settings Page**

First, you need to know how to access your wireless router's settings. Usually you can do this by typing in "192.168.1.1" into your web browser, and then enter the correct user name and password for the router. This is different for each router, so first check your router's user manual.

You can also use Google to find the manuals for most routers online in case you lost the printed manual that came with your router purchase. For your reference, here are direct links to the manufacturer's site of some popular router brands – Linksys, Cisco, Netgear, Apple AirPort, SMC, D-Link, Buffalo, TP-LINK, 3C0m, Belkin.

**Create a unique password on your router**

Once you have logged into your router, the first thing you should do to secure your network is to change the default password* of the router to something more secure.

This will prevent others from accessing the router and you can easily maintain the security settings that you want. You can change the password from the Administration settings on your router's settings page. The default values are generally admin / password.

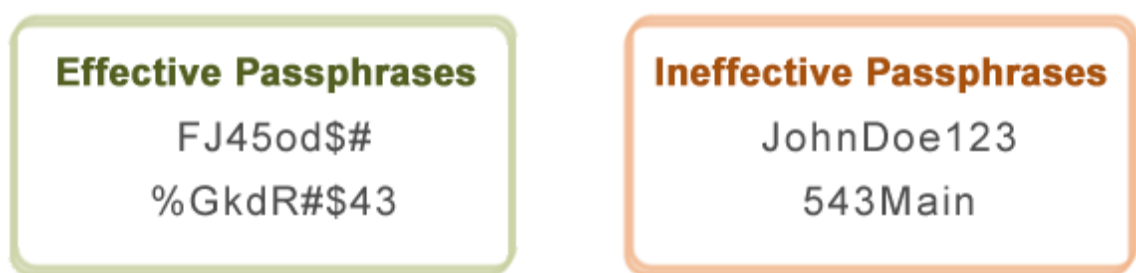

Fig.1 Sample Password

**[*] What do the bad guys use** - This is a public database [1] of default usernames and passwords of wireless routers, modems, switches and other networking equipment. For instance, anyone can easily make out from the database that the factory-default settings for Linksys equipment can be accessed by using admin for both username and password fields.

**Change your Network's SSID name**

The SSID (or Wireless Network Name) of your Wireless Router is usually pre-defined as "default" or is set as the brand name of the router (e.g., Linksys). Although this will not make your network inherently* more secure, changing the SSID name of your network is a good idea as it will make it more obvious for others to know which network they are connecting to.

This setting is usually under the basic wireless settings in your router's settings page. Once this is set, you will always be sure that you are connecting to the correct Wireless network even if there are multiple wireless networks in your area. Don't use your name, home address or other personal information in the SSID name.
Hide your network ID
A router broadcasts its SSID to anyone within range. You can alter the router settings to not broadcast the SSID and therefore avoid alerting hackers to the network's existence.
 **[*] What do the bad guys use** - Wi-Fi scanning tools like in SSID [2a] (Windows) and Kismet[2b](Mac, Linux) are free and they will allow anyone to find all the available Wireless Networks in an area even if the routers are not broadcasting their SSID name.

**Enable Network Encryption**

In order to prevent other computers in the area from using your internet connection, you need to encrypt your wireless signals.

There are several encryption methods for wireless settings, including WEP, WPA (WPA-Personal), and WPA2 (Wi-Fi Protected Access version 2). WEP is basic encryption and therefore least secure (i.e., it can be easily cracked*, but is compatible with a wide range of devices including older hardware, whereas WPA2 is the most secure but is only compatible with hardware manufactured since 2006.

| Manufacturer | Model | Version | Access Type | Username | Password | Notes |
|---|---|---|---|---|---|---|
| Cabletron | Netgear modem/router and SSR | | | netman | (none) | |
| NetGear | RM356 | None | Telnet | (none) | 1234 | shutdown the router via internet |
| Netgear | MR-314 | 3.26 | HTTP | admin | 1234 | |
| Netgear | RT314 | | HTTP | admin | admin | |
| Netgear | RP614 | | HTTP | admin | password | |
| Netgear | RP114 | 3.26 | Telnet | (none) | 1234 | telnet 192.168.0.1 |
| Netgear | WG602 | Firmware Version 1.04.0 | HTTP | super | 5777364 | |
| Netgear | WG602 | Firmware Version 1.7.14 | HTTP | superman | 21241036 | |
| Netgear | WG602 | Firmware Version 1.5.67 | HTTP | super | 5777364 | |
| Netgear | MR814 | | HTTP | admin | password | |

Fig.2 Default Configuration List for NetGear Wireless Routers

In above figure we can see the default configuration of routers made by NetGear Company. These routers are most widely used in wireless networks. Following settings should always been changed from default one to prevent hackers from exploiting the devices.
1. Encryption settings.
2. SSID broadcast setting
3. Default Password
4. Use HTTPS instead of HTTP
5. Session Logging

To enable encryption on your Wireless network, open the wireless security settings on your router's configuration page. This will usually let you select which security method you wish to choose; if you have older devices, choose WEP, otherwise go with WPA2. Enter a passphrase to access the network; make sure to set this to something that would be difficult for others to guess, and consider using a combination of letters, numbers, and special characters in the passphrase.

[*] **What do the bad guys use** - Air Crack and coWPAtty are some free tools that allow even non-hackers to crack the WEP / WPA (PSK) keys using dictionary or brute force techniques. A video on YouTube suggests that Air Crack may be easily used to break Wi-Fi encryption using a jail-broken iPhone or an iPod Touch.

**Filter MAC addresses**

Whether you have a laptop or a Wi-Fi enabled mobile phone, all your wireless devices have a unique MAC address (this has nothing to do with an Apple Mac) just like every computer connected to the Internet has a unique IP address. For an added layer of protection, you can add the MAC addresses of all your devices to your wireless router's settings so that only the specified devices can connect to your Wi-Fi network.

MAC addresses are hard-coded into your networking equipment, so one address will only let that one device on the network. It is, unfortunately, **possible to spoof a MAC address\***, but an attacker must first know one of the MAC addresses of the computers that are connected to your Wireless network before he can attempt spoofing.

To enable MAC address filtering, first make a list of all your hardware devices that you want to connect to your wireless network\*\*. Find their MAC addresses, and then add them to the MAC address filtering in your router's administrative settings. You can find the MAC address for your computers by opening Command Prompt and typing in "ipconfig /all", which will show your MAC address beside the name "Physical Address". You can find the MAC addresses of Wireless mobile phones and other portable devices under their network settings, though this will vary for each device.

[*] **What do the bad guys use** - Someone can change the MAC address of his or her own computer and can easily connect to your network since your network allows connection from devices that have that particular MAC address. Anyone can determine the MAC address of your device wireless using a sniffing tool like Nmap and he can then change the MAC address of his own computer using another free tool like MAC Shift.

**Reduce the Range of the Wireless Signal**

If your wireless router has a high range but you are staying in a small studio apartment, you can consider decreasing the signal range by either changing the mode of your router to 802.11g (instead of 802.11n or 802.11b) or use a different wireless channel.

You can also try placing the router under the bed, inside a shoe box or wrap a foil around the router antennas so that you can somewhat restrict the direction of signals.

**Apply the Anti-Wi-Fi Paint** – Researchers have developed a special Wi-Fi blocking paint that can help you stop neighbors from accessing your home network without you having to set up encryption at the router level. The paint contains chemicals that blocks radio signals by absorbing them. "By coating an entire room, Wi-Fi signals can't get in and, crucially, can't get out."

**Upgrade your Router's firmware**

You should check the manufacturer's site occasionally to make sure that your router is running the latest firmware. You can find the existing firmware version of your router using from the router's dashboard at 192.168.*.

**Connect to your Secure Wireless Network**

To conclude, MAC Address filtering with WPA2 (AES) encryption (and a really complex passphrase) is probably the best way to secure your wireless network.

Once you have enabled the various security settings in your wireless router, you need to add the new settings to your computers and other wireless devices so that they all can connect to the Wi-Fi network. You can select to have your computer automatically connect to this network, so you won't have to enter the SSID, passphrase and other information every time you connect to the Internet.

**Monitor your network for intruders.**

You should always make sure you have an eye on what's going on, that you are tracking attack trends. The more you know about what malicious security crackers are trying to do to your network, the better the job of defending against them you can do. Collect logs on scans and access attempts, use any of the hundreds of statistics generating tools that exist to turn those logs into more useful information, and set up your logging server to email you when something really anomalous happens. As a certain cartoon military SpecOps team from the 1980s would tell you, knowing about the danger is half the battle.

**Check that your device does not auto-connect to Wi-Fi signals**

If your device is set to automatically connect to available Wi-Fi networks, then you run the risk of automatically connecting to unknown and potentially dangerous networks. You should switch off auto-connect on your device settings page – refer to the manufacturer's instructions for more details.

**Security browser extensions**

One essential browser extension I would recommend is HTTPS Everywhere from the Electronic Frontier Foundation (EFF). This allows you to have a secure connection when you visit common sites like Google, Yahoo, ebay, Amazon, and more. It also allows you to create your own XML config file to add more sites not listed. It's available for both Chrome and Firefox and works with Windows, Mac, and Linux.

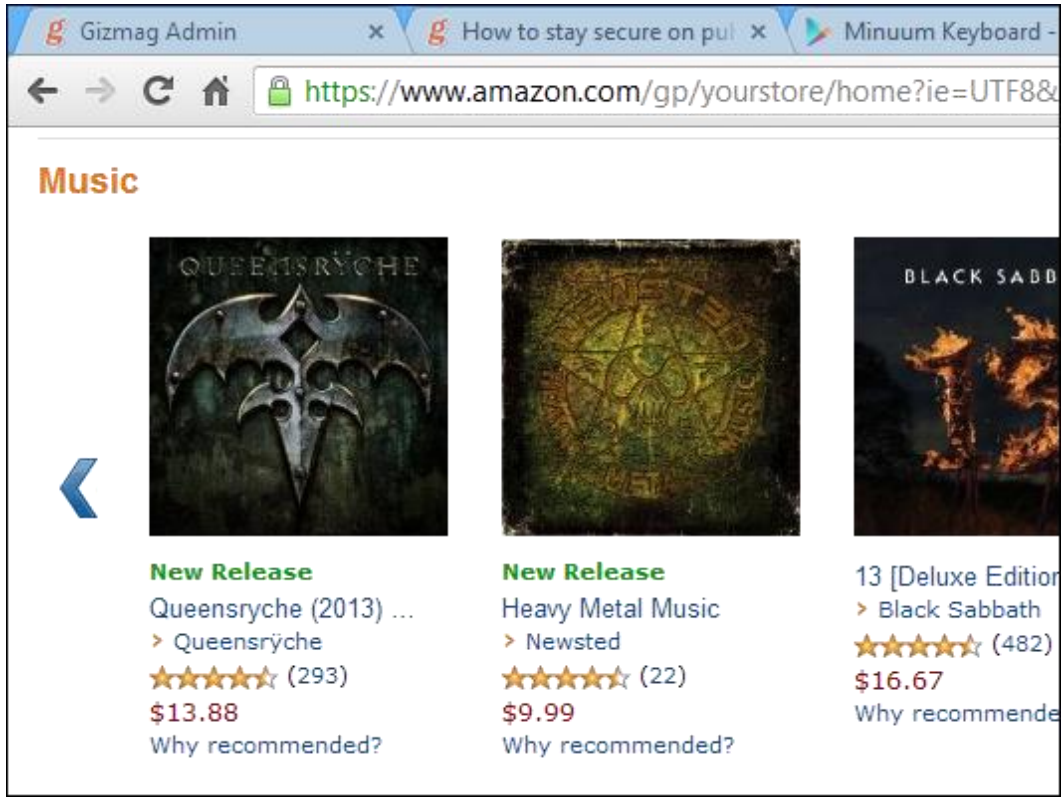

Most of the security tips one can offer about wireless networking are the sort of thing someone might call "common sense". Unfortunately, there's an awful lot of "common sense" floating around out there, and it's not easy to keep it all in mind all the time. You should always check up on your wireless networks and mobile computers regularly to make sure you aren't missing something important, and you should always double-check your assumptions to make sure you aren't wasting your energy on something not only unnecessary, but entirely useless, when more effective security measures could use your attention.

IV. CONCLUSION

Your wireless network will now be a lot more secure and intruders may have a tough time intercepting your Wi-Fi signals. Wi-Fi gives freedom to users so they can access network easily but at the same time it gives easy way to the hackers to penetrate and exploit the network if proper security is not there. Hackers can stay anonymous on wireless network with less effort and can easily target their victims by using various methods such as sniffing, dummy WAP etc. In order to prevent such intrusion and detect such attacks we should always follow best practices such as changing default configuration, using Wireless intrusion detection systems and so on. If proper security is followed then security professionals can not only stop hackers from exploiting the network but also catch them by hacking the hacker.